\documentclass[final]{svjour3b}

\usepackage{graphicx}
\usepackage{amsmath}
\usepackage{amsfonts,amssymb}
\usepackage{cite}
\usepackage{mathptmx}
\usepackage[colorlinks=true,linkcolor=blue,citecolor=blue,pagecolor=black,menucolor=black,raiselinks=false]{hyperref}

\newcommand{\fig}[1]{Fig.~\ref{#1}}

\newcommand{\eq}[1]{Eq.~(\ref{#1})}
 \newcommand{\ub}{{\rm ub}}
\newcommand{\bd}{{\rm b}} \newcommand{\piad}{\pi_{\rm ad}}

\begin{document}

\title{Transport by molecular motors in the presence of static
  defects}

\author{Yan Chai \and Reinhard Lipowsky \and Stefan Klumpp}

\institute{
  Yan Chai and Reinhard Lipowsky \at Max Planck Institute of Colloids and Interfaces, Science Park Golm, 14424 Potsdam, Germany\\
  \email{chai@mpikg.mpg.de} \and Stefan Klumpp \at Center for
  Theoretical Biological Physics and Department of Physics, 
  University of California at San Diego,
  La Jolla, CA 92093-0374, USA
}

\maketitle


\begin{abstract}

  The transport by molecular motors along cytoskeletal filaments is
  studied theoretically in the presence of static defects. The
  movements of single motors are described as biased random walks
  along the filament as well as binding to and unbinding from the
  filament. Three basic types of defects are distinguished, which
  differ from normal filament sites only in one of the motors'
  transition probabilities. Both stepping defects with a reduced
  probability for forward steps and unbinding defects with an
  increased probability for motor unbinding strongly reduce the
  velocities and the run lengths of the motors with increasing defect
  density. For transport by single motors, binding defects with a
  reduced probability for motor binding have a relatively small effect
  on the transport properties. For cargo transport by motors teams,
  binding defects also change the effective unbinding rate of the
  cargo particles and are expected to have a stronger effect.


\end{abstract}


\section{Introduction}


The interior of living cells is characterized by highly organized
complex structures. To build and maintain these internal structures,
cells rely on directed active transport of various types of cargoes to
different destinations within the cell. This transport is driven by
molecular motors which use the energy derived from the hydrolysis of
adenosine triphosphate (ATP) to move along cytoskeletal filaments
\cite{Howard2001, Schliwa2003}. There are three large families of
cytoskeletal motors, kinesins and dyneins, which move along
microtubules, and myosins, which move along actin filaments
\cite{Howard2001, Schliwa2003}.

Since cells provide crowded environments, motors moving along
filaments encounter a variety of other molecules bound to the same
filaments, which may hinder their movement. These obstacles may
represent other motors of the same type, and the traffic phenomena
that arise in such systems with many motors have been studied
extensively in recent years. Many theoretical studies have explored
the formation of traffic jams and non-equilibrium phase transitions
\cite{Lipowsky_Nieuwenhuizen2001,Klumpp_Lipowsky2003,Parmeggiani_Frey2003,Evans_Santen2003,Klein_Juelicher2005,Nishinari_Chowdhury2005},
and traffic jams of molecular motors have recently been observed in
several experimental studies
\cite{Konzack_Fischer2005,Nishinari_Chowdhury2005,Leduc_Prost2004}. A
system with two different species of motors that move into opposite
direction has also been studied theoretically and is predicted to
exhibit spontaneous symmetry breaking and the formation of separate
traffic lanes for the two directions \cite{Klumpp_Lipowsky2004}.

In addition to molecular motors, a variety of other molecules can bind
to filaments and affect the movement of these motors.  An important
example is given by microtubule-associated proteins (MAPs), which bind
to microtubules to control their structure and stability. In addition,
MAPs can modulate the movement of the motors along the microtubules.
When overexpressed {\it in vivo}
\cite{SatoHarada_Hirokawa1996,Bulinski_Sheetz1997,Ebneth_Mandelkow1998}
or added to microtubule gliding assays with kinesin or dynein motors
{\it in vitro}
\cite{Heins_Mandelkow1991,Paschal_Vallee1989,Lopez_Sheetz1993,Hagiwara_Hirokawa1994},
MAPs decrease or completely inhibit motility of motors. More recent
experiments using lower concentrations of MAPs and tracking the
movements of individual motors show that most MAPs studied so far
affect motor movements by modulating binding of the motors to
microtubules. For example, the tau protein, a MAP specific for
neurons, has been shown to decrease the binding of kinesin and dynein
motors to microtubules
\cite{Seitz_Mandelkow_2002,Vershinin_Gross2007,Dixit_Holzbaur2008}.
Its effect depends on the tau isoform \cite{Vershinin_Gross2007}, is
more pronounced for kinesin than for dynein motors
\cite{Ebneth_Mandelkow1998,Dixit_Holzbaur2008,Vershinin_Gross2008},
and has a stronger effect on cargoes pulled by several motors than on
individual motors, see Refs.
\cite{Seitz_Mandelkow_2002,Vershinin_Gross2007,Klumpp_LipowskyCoopTr}
and the discusson below. These subtle and highly specific effects seen
at low tau concentration
\cite{Seitz_Mandelkow_2002,Vershinin_Gross2007,Dixit_Holzbaur2008}
suggest that tau (and other MAPs) may play important roles as
regulators of transport in cells, and function as general transport
inhibitors only under pathological conditions \cite{Mandelkow_2004}.
For example, the differential effects on kinesins and dynein suggest
that tau can control the direction of motion of cargoes that are
carried by both types of motors, as discussed in Ref.
\cite{Mueller_Lipowsky2008}.

When modeling large-scale transport by molecular motors, static
molecules bound to the filaments can be considered as local properties
of the filament. They represent static or quenched defects of the
filament that affect the motor dynamics locally. The same theoretical
description may then be used for other types of defects that cause
local effects on motor transport. Such defects may for example be
local modifications of the filament themselves such as microtubule
lattice defects or a variety of post-transcriptional modifications of
tubulin, the subunit of microtubules. Some of these modifications have
been shown to affect the microtubule-binding or the movement of motors
\cite{Larcher_Denoulet_1996,Reed_Verhey_2006, Lakamper_Meyhofer2005}.
Finally, in addition to these naturally occurring defects, artificial
'roadblocks' such as inactive motor mutants have been used in several
experiments to perturb the movement of active motors in order to study
the mechanisms of motor function
\cite{Crevel_Cross2004,Seitz_Surrey2006}.

In this paper, we study the effects of various types of defects on the
movements of molecular motors using the lattice model introduced in
Ref. \cite{Lipowsky_Nieuwenhuizen2001}. Here we use the simple
description of the dynamics of motor stepping provided by the lattice
model to distinguish three basic types of static defects and to study
their effects on single motors as well as on the motor traffic in
many-motor systems. The three basic types of defects are given by
filament sites that differ from the other filament sites in one of
three motor parameters: (i) {\em stepping} defects have an altered
forward stepping probability, (ii) {\em unbinding} defects have an
altered unbinding probability, and (iii) {\em binding} defects have an
altered binding probability. Some cases that have been studied
previously can be considered as special cases of this general
approach. For example a single stepping defect has been studied in
Ref. \cite{Pierobon_Frey2006}, and a single unbinding defect without
unbinding from non-defect sites in Ref. \cite{Mirin_Kolomeisky2003}.
Very recently, binding defects have been studied in Ref.
\cite{Grzeschik_Santen2008}. Stepping defects have also been
investigated extensively for one-dimensional exclusion processes
\cite{Janowsky_Lebowitz1992,Tripathy_Barma1997,Kolomeisky1998,Klumpp_Lipowsky_PRE2004},
which, in our model, correspond to movement of motors along a filament
without binding and unbinding. We also note that in the statistical mechanics literature such 
defects are
classified as 'sitewise' disorder, since the anomalous properties are
related to a fraction of the lattice sites, as opposed to the case of 'particlewise'
disorder, for which some of the moving particles exhibit anomalous
properties \cite{Krug2000}.

The paper is organized as follows: In section \ref{Lattice model with
  different types of defects}, we introduce the lattice model and the
system geometry used in this study as well as the description and
classification of defects. We discuss the modeling of known biological
defects such as MAPs within this model. We then study stepping defects
in section \ref{Transport by molecular motors in the presence of
  stepping defects}, unbinding defects in section \ref{Transport by
  molecular motors in the presence of unbinding defects} and binding
defects in section \ref{Transport by molecular motors in the presence
  of binding defects}. We conclude with a few general remarks on the
use of defects in transport.


\section{Lattice model with different types of defects}
\label{Lattice model with different types of defects}


\subsection{Lattice model for the traffic of molecular motors}

To study the effects of various types of defects and the transport by
molecular motors, we extend the lattice model introduced in Ref.\
\cite{Lipowsky_Nieuwenhuizen2001}, which we have previously used to
describe both the movement of single motors
\cite{Lipowsky_Nieuwenhuizen2001,Nieuwenhuizen_Lipowsky2004,Klumpp_LipowskyAct}
as well as the traffic in many-motor systems
\cite{Lipowsky_Nieuwenhuizen2001,Klumpp_Lipowsky2003}. This model
describes the movements of a single molecular motor along a filament
as a random walk on a (generally three-dimensional) lattice, which
contains one or several lines of lattice sites that represent
filaments. The lattice constant $\ell$ is given by the step size of a
motor moving actively along a filament. Per unit time $\tau$, a motor
at a filament site makes a forward step along the filament with
probability $\alpha$, unbinds to each of the four neighboring
non-filament sites with probability $\epsilon/6$, and remains at the
same site with probability $\gamma=1-\alpha-4\epsilon/6$. Motors at
non-filament sites perform symmetric random walks and move to each
nearest neighbor site with probability $1/6$. The choice of this
probability implies that the time scale $\tau$ is given by the
diffusion coefficient of unbound motors $D_\ub$ as
$\tau=\ell^2/D_\ub$. If an unbound motor moves to a filament site, it
binds to it with the sticking probability $\piad$. If $\piad\neq1$,
this condition modifies the probability for the movement from a
non-filament site to a filament site to $\piad/6$.  In general, we can
model both freely suspended filaments, for which the filament site is
connected to four neighboring non-filament sites, and immobilized
filaments, for which the number of nearest neighbors is at most equal
to three. In the simulations reported below, we focused on freely
suspended filaments.

In addition to the dynamics of single motors, the lattice model can
also describe systems with many interacting motors. In the simplest
case, these motors interact only through their mutual exclusion from
lattice sites, which is implemented in the model by not allowing any
steps to sites that are occupied by another motor.  Typically the
density of motors at non-filament sites is much lower than at filament
site, so that the exclusion rule affects mainly binding to the
filament and movement along it. By virtue of this exclusion rule, our
model is a variant of driven lattice gas models or exclusion
processes, which have been studied extensively as model systems for
transport processes and non-equilibrium phase transitions
\cite{Schmittmann_Zia1995,Schuetz2001}.

\begin{figure}
  \centering
  \includegraphics[width=11cm,clip]{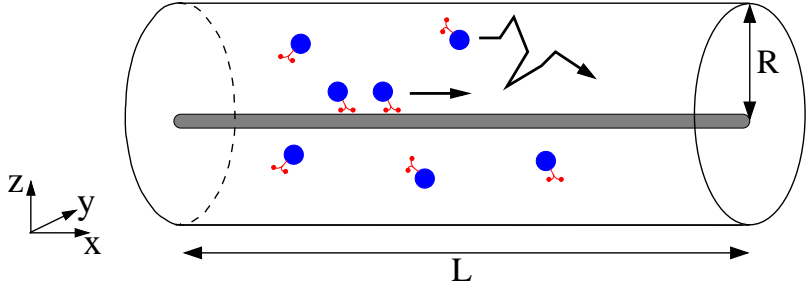}
  \caption{Molecular motors inside a cylindrical tube with a filament
    aligned along its axis. This tube system mimics the geometry of
    elongated cellular structures such as axons. The tube has the
    length $L$ and the radius $R$. Motors bound to the filament move
    actively along the filament in a directed fashion, while unbound
    motors perform diffusive movements. The boundary condition is
    periodic along the $x$-axis.}
  \label{tube system}
\end{figure}

Throughout this article, we will study systems that have a tube-like
geometry as shown in \fig{tube system}. In these systems a single
filament is located on the axis of a cylindrical tube with length $L$
and radius $R$. This geometry mimics the structure of some types of
cells, such as axons of nerve cells or hyphae of fungi, which are
approximately tubular and have a unidirectional microtubule
cytoskeleton \cite{Goldstein_Yang2000}. Similar tube-like systems have
previously been studied with various types of boundary conditions
\cite{Lipowsky_Nieuwenhuizen2001,Klumpp_Lipowsky2003,Klumpp_Lipowsky2005,Mueller_Lipowsky2005}.
In order to keep the discussion simple, we will focus on periodic
boundary conditions in the following.

For periodic boundary conditions, the case without defects is particularly simple and has
been solved exactly \cite{Klumpp_Lipowsky2003}. In this case, the densities of bound and unbound motors, $\rho_\bd$
and $\rho_\ub$, respectively, are spatially homogeneous and satisfy a
binding-unbinding balance condition,
\begin{equation}
  \pi_{\rm ad} \rho_{\rm ub} (1-\rho_\bd) = \epsilon  \rho_{\rm b} (1-\rho_\ub)\approx\epsilon\rho_\bd,
\end{equation}
and the motor current $J$ along the filament is given by $J=\alpha
\rho_\bd (1-\rho_\bd)$.  In the case of a single motor, the balance
equation is
\begin{equation}
  \pi_{\rm ad} \rho_{\rm ub}  = \epsilon  \rho_{\rm b},
  \label{balance}
\end{equation}
from which one can derive the steady-state probability that the motor
is bound to the filament, $P_\bd=\sum_x\rho_\bd = \rho_\bd L/\ell$,
which is given by
\begin{equation}
  P_{\rm  b} =\frac{\pi_{\rm ad}}{\pi_{\rm ad}+\epsilon N_{\rm ch}},
\end{equation}
where $N_{\rm ch}$ is the number of unbound channels, i.e., the number
of lines of lattice sites parallel to the filament in a discretized
tube with cross section $\phi=(1+N_{\rm ch})\approx\pi R^2$ for
sufficiently large radius $R$. The effective motor velocity, averaged
over the bound and unbound states of the motor, is then obtained as
\begin{equation}
  v_{\rm eff}=P_{\rm  b} v_{\rm b}=
  \frac{\alpha \pi_{\rm ad}}{\pi_{\rm ad}+\epsilon N_{\rm ch}}\frac{\ell}{\tau},
  \label{v_normal}
\end{equation}
where $v_{\rm b}=\alpha\ell/\tau$ is the velocity of the bound motor.

\subsection{Lattice model with different types of defects}

\begin{figure}
  \centering
  \includegraphics[width=11cm,clip]{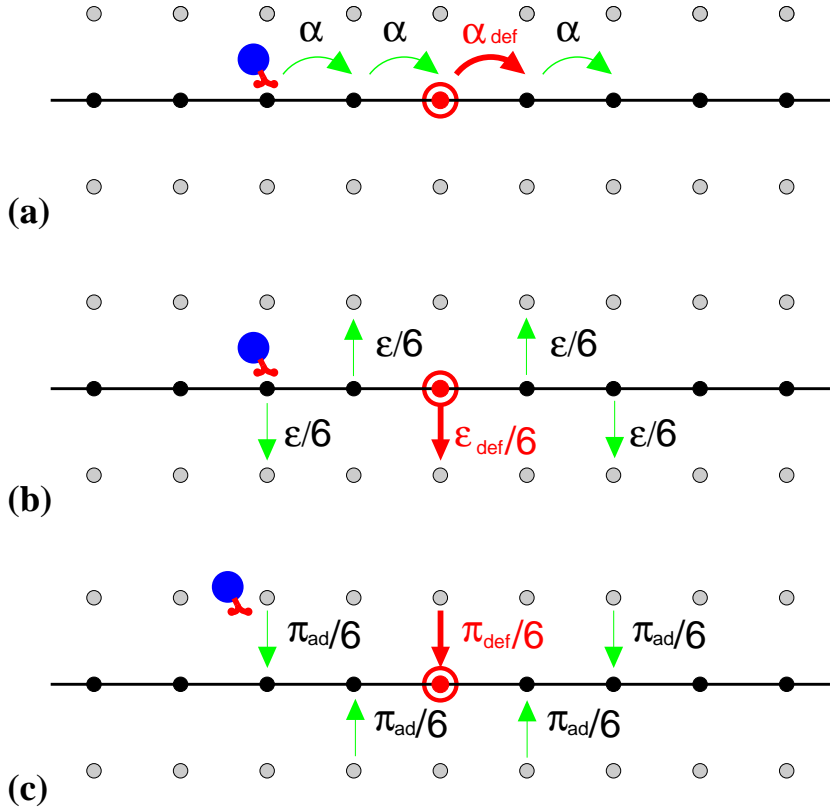}
  \caption{Three types of defects studied in this article: (a)
    Stepping defect indicated by an encircled site with a modified
    forward stepping probability $\alpha_{\rm def}$ compared to other
    filament sites; (b) Unbinding defect with a modified unbinding
    probability $\epsilon_{\rm def}$; and (c) Binding defect with a
    modified binding probability $\pi_{\rm def}$. All other transition
    probabilities are the same at the defect site as at the other
    filament sites.}
  \label{model of defects}
\end{figure}

Inhomogeneities of the filament such as those mentioned in the
introduction may affect one or several of the motor properties. This
can be described within the lattice model by modifying one or several
of the hopping probabilities compared to the homogeneous situation. In
the following, we distinguish three basic types of defects which are
characterized by a {\it single} parameter that differs from the
homogeneous case as show in Fig. \ref{model of defects}: (i) Stepping
defects have a changed probability $\alpha_{\rm def}$ for forward
movement, but unchanged binding probability $\pi_{\rm ad}$ and
unbinding probability $\epsilon$; (ii) unbinding defects have a
changed unbinding probability $\epsilon_{\rm def}$; and (iii) binding
defects have a modified sticking probability $\pi_{\rm def}$. In all
three cases, the dwell probability $\gamma_{\rm def}$ also needs to be
adapted, so that the sum of all probabilities is again equal to one.

More complicated types of defects can be considered as combinations of
these basic defects. For example an inaccessible site due to a large
immobile protein bound to the filament, such as an inactive mutant
motor, can be described by a combination of a stepping defect and a
binding defect with $\alpha_{\rm def}=\pi_{\rm def}=0$.

\begin{table}
  \caption{Overview of different types of defects and their effects on molecular motor movement}
  \begin{center}
    \begin{tabular}{p{40pt}|c||c|c|c||l|c}
      \hline
      defects  & motors & stepping & unbinding & binding & references & note\\
      \hline
      \hline
      tau & kinesin & -- & -- & reduced & \cite{Seitz_Mandelkow_2002,Vershinin_Gross2007,Dixit_Holzbaur2008}& a \\\hline
      tau & kinesin & reduced & increased & reduced & \cite{Dixit_Holzbaur2008} & b \\\hline
      MAP2c & kinesin & -- & -- & reduced & \cite{Seitz_Mandelkow_2002} & \\\hline
      MAP4 & kinesin & reduced & -- & -- & \cite{Tokuraku_Kotani2007} &  c\\\hline
      tau & dynein-dynactin & reduced & increased & reduced & \cite{Dixit_Holzbaur2008} & b \\\hline
      tau & dynein & --  & -- & reduced & \cite{Vershinin_Gross2008} &  \\\hline
      \hline
      de-acetylated $\alpha$-tubulin& kinesin & reduced  & -- & reduced & \cite{Reed_Verhey_2006} & \\\hline
      tubulin without E-hook & kinesin & reduced  & -- & -- & \cite{Lakamper_Meyhofer2005}& \\\hline
      \hline
      inactive motor mutant & kinesin & reduced to 0 & ? & reduced to 0 & \cite{Crevel_Cross2004,Seitz_Surrey2006} & d\\\hline
    \end{tabular}
    \newline
    \begin{tabular}{l p{300pt}}
      a & Refs.\ \cite{Vershinin_Gross2007,Dixit_Holzbaur2008} also report shorter run lengths, i.e.\ increased unbinding, not observed in Ref. \cite{Seitz_Mandelkow_2002}. In addition, Ref. \cite{Dixit_Holzbaur2008} reports also a substantial fraction of immobile motors bound to microtubules.\\
      b &  Under the conditions of this experiment motors have rather long run lengths in the absence of tau, longer than in the experiments of Ref. \cite{Seitz_Mandelkow_2002,Vershinin_Gross2007,Vershinin_Gross2008}.\\
      c & The effect depends on which isoform of MAP4 is used. A 5-repeat isoform exhibits a strong effect, while the other isoforms studied showed only small effects \cite{Tokuraku_Kotani2007}.\\
      d & Refs.\  \cite{Crevel_Cross2004,Seitz_Surrey2006} report conflicting results for the effect on unbinding.
    \end{tabular}
  \end{center}
  \label{Examples of defects and their effect on motor properties}
\end{table}

Table \ref{Examples of defects and their effect on motor properties}
lists examples of defects that have been characterized experimentally
and summarizes their effects on molecular motors. MAPs such as the tau proteins
essentially represent binding defects (an exception is MAP4
\cite{Tokuraku_Kotani2007}, see Table \ref{Examples of defects and
  their effect on motor properties}). They reduce the binding rate of
kinesin to microtubules and have no or only a weak effect on the
velocity of bound kinesins as well as on the unbinding rate or the run
length, the distance moved along the filament before unbinding
\cite{Seitz_Mandelkow_2002,Vershinin_Gross2007}. Similar effects have
been observed for dynein motors \cite{Dixit_Holzbaur2008}, which in
general are less affected by MAPs than kinesins. {\it In vivo}, tau
has also been shown to reduce the run length for vesicular cargoes.
The increase in unbinding rate for these cargoes is most likely a
consequence of the fact that these cargoes are pulled by several
motors rather than a single motor, since for cargoes pulled by several
motors, the unbinding rate is a function of the single-motor binding
rate \cite{Seitz_Mandelkow_2002,Klumpp_LipowskyCoopTr}. This effect
has been demonstrated {\it in vitro} for tau and beads pulled by
several kinesins \cite{Vershinin_Gross2007}.  Therefore defects that
are binding defects for individual motors, can be both binding and
unbinding defects for cargoes pulled by multiple motors.

The effects of post-translational modifications of microtubules on
motor movements have not been characterized in much mechanistic
detail. One case for which the effect is known is
acetylation/deacetylation of a particular lysine residue of
$\alpha$-tubulin for which it has been shown that kinesin binds more
strongly to the acetylated form than to the de-acetylated form
\cite{Reed_Verhey_2006}. Microtubules containing de-actelyated tubulin
subunits therefore provide another example of binding defects.
Microtubule lattice defects are believed to cause unbinding of motors,
see, e.g. Ref. \cite{Davis_Gross2002}, and would thus represent
unbinding defects. While this scenario is plausible, it has not been
studied systematically and there is no direct experimental evidence
for it.

Finally, the artificial 'roadblock' motor mutants used in Refs.
\cite{Crevel_Cross2004,Seitz_Surrey2006} represent blocked sites, i.e.
combinations of a stepping defect with very low, essentially zero,
stepping rate and a binding defect. Whether they also affect unbinding
is unclear, since the two experiments in Refs.
\cite{Crevel_Cross2004,Seitz_Surrey2006} reported conflicting results,
for a discussion see also Ref. \cite{Klumpp_Lipowsky2008}. In another
recent experiment, some motors were inactivated by irreversible
crosslinking to a microtubule to obtain blocked sites
\cite{Dreblow_Boehm2008}.


\section{Transport by molecular motors in the presence of stepping
  defects}
\label{Transport by molecular motors in the presence of stepping
  defects}


\subsection{Single motor with stepping defects}

We start by considering stepping defects. At a stepping defect, the
motor has the forward stepping probability $\alpha_{\rm def}$, while
the binding and unbinding parameters are the same as at the other
filament sites. We note that the probability to remain at the site is
also changed compared to other sites and given by $\gamma_{\rm
  def}=1-\alpha_{\rm def}-4\epsilon /6$. We consider a single filament
in a tube as shown in Fig. \ref{tube system} with a density $\rho_{\rm
  def}$ of stepping defect sites. To keep the discussion simple we
study the case where the defect sites are arranged periodically on the
filament. This situation is then equivalent to a system with a single
defect, length $L=\ell/\rho_{\rm def}$ and periodic boundary
conditions.

First, we consider the effect of stepping defects on a single motor.
As the stepping defects do not affect the binding and unbinding
probabilities of the motor, one may expect that the binding
probability is the same as in the absence of defects. However,
relation \eqref{balance}, which describes a local balance of binding
and unbinding, is not valid in the presence of stepping defects, since
the motor densities are not constant along the filament because of the
prolonged waiting times of the motor at the defect sites. Thus
relation \eqref{balance} has to be replaced by the global balance of
binding and unbinding which remains valid if the densities are not
constant and is given by
\begin{equation}
  \sum_{x}\pi_{\rm ad}\rho_{\rm ub}(x,y_{\rm nn},z_{\rm nn})= \sum_{x} \epsilon \rho_{\rm b} (x),
  \label{balance_global}
\end{equation}
where $y_{\rm nn}$ and $z_{\rm nn}$ are the perpendicular coordinates
of a single channel of non-filament sites that are the nearest
neighbors of the filament sites, e.g. $y_{\rm nn}=\ell$ and $z_{\rm
  nn}=0$. The inhomogeneity of the unbound density is relatively small
because the fast motor diffusion tends to smooth the unbound density
profile and taking the unbound density to be independent of the
coordinates perpendicular to the filament, i.e. $\rho_{\rm ub} (x,y,z)
\simeq \rho_{\rm ub}(x)$, is usually a very good approximation
\cite{Klumpp_Lipowsky2003,Klumpp_Lipowsky2005}. Within this
"two-state" approximation, the probabilities $P_{\rm b}$ and $P_{\rm
  ub}$ to find the motors in a bound or an unbound state,
respectively, are given by
\begin{equation}
  P_{\rm  b}=\sum_{x}\rho_{\rm b} (x)  \hspace{10pt} {\rm and} \hspace{10pt} P_{\rm ub}=N_{\rm ch}\sum_{x}\rho_{\rm ub}(x),
  \label{bound and unbound probabilities}
\end{equation}
and satisfy the normalization condition
\begin{equation}
  P_{\rm b}+P_{\rm ub}=P_{\rm b}+N_{\rm ch}\sum_{x}\rho_{\rm ub}(x)=1.
  \label{normalization condition for bound and unbound motors}
\end{equation}
The flux balance relation \eqref{balance_global} then becomes
\begin{equation}
  \pi_{\rm ad}\frac{P_{\rm ub}}{N_{\rm ch}}=\pi_{\rm ad}\frac{1-P_{\rm b}}{N_{\rm ch}}=\epsilon P_{\rm b},
  \label{flux balance for two-state approximation}
\end{equation}
which leads to
\begin{equation}
  P_{\rm  b}=\frac{\pi_{\rm ad}}{\pi_{\rm ad}+\epsilon N_{\rm ch}},
  \label{total bound probability for single motor with stepping defects}
\end{equation}
the same expression as for the case without stepping defects.

To obtain the effective velocity of the motor, we introduce an
effective passing time to describe the movement of the bound motor. We
assume that the motor spends the time $\tau_0=\tau/\alpha$ at a normal
site and the time $\tau_{\rm def}$ at a defect site.  Since there are
$L/\ell-1$ normal sites and only one defect site on a filament segment
of length $L$, the total time to move through such segment is $t_{\rm
  tot}=(L-\ell)\tau/\ell\alpha+\tau_{\rm def}$, provided the motor
typically remains bound to the filament during such a run.  The
velocity of a bound motor can be estimated by
\begin{equation}
  v_{\rm b,eff}=\frac{L}{t_{\rm tot}} =
  \frac{\alpha}{1-\frac{\ell}{L}(1-\frac{\alpha\tau_{\rm def}}{\tau})}\frac{\ell}{\tau}. \label{current from the total passing time}
\end{equation}
The effective velocity, which characterizes the motor movement
including the diffusive excursions upon unbinding, is then given by
$v_{\rm eff}=v_{\rm b,eff} P_\bd$.

The time $\tau_{\rm def}$ to pass a defect remains to be specified. In
the limit of a sufficiently weak defect and sufficiently processive
motors as assumed so far, this time is given by the inverse of the
defect stepping probability, i.e. $\tau_{\rm def}=\tau/\alpha_{\rm
  def}$. In general, however, there are two ways in which a motor can
pass a stepping defect: the motor can either slowly step through the
defect along the filament or it may unbind from the filament and
rebind to it after diffusing around the defect. The relative
importance of these two pathways depends on their relative
probabilities: when $\alpha_{\rm def}\gg\epsilon$, the direct path
through the defect dominates, while unbinding and diffusion will be
the dominant pathway for $\alpha_{\rm def}\ll\epsilon$.

\begin{figure}
  \centering
  \includegraphics[width=12cm,clip]{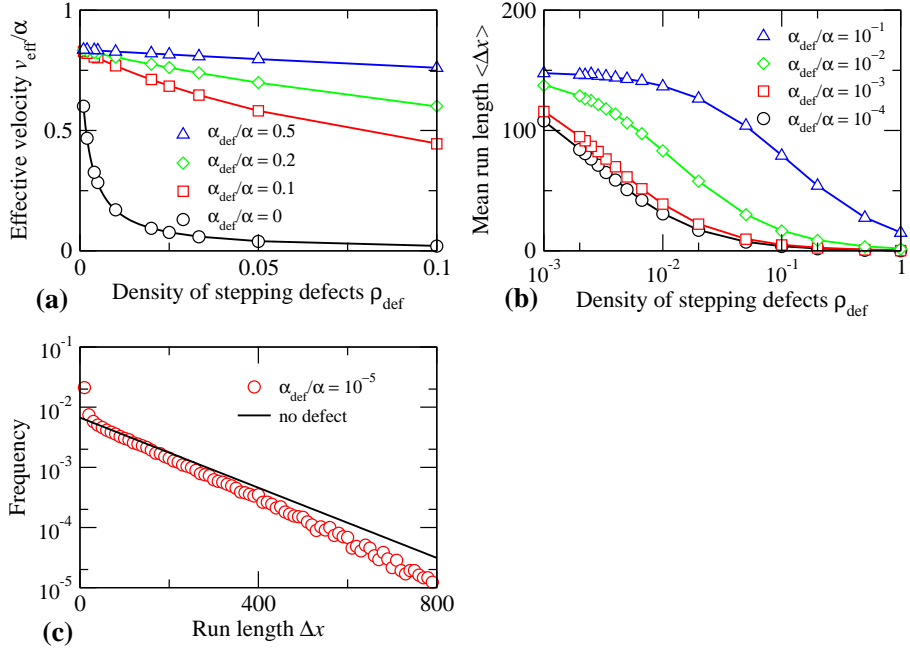}
  \caption{Transport properties of single motors in the presence of
    stepping defects: (a) Normalized effective velocity $v_{\rm
      eff}/\alpha$ (in units of $\ell/\tau$) and (b) Average run
    length $\langle\Delta x\rangle$ (in units of $\ell$) as a function
    of the defect density $\rho_{\rm def}$ for different values of the
    defect strength $\alpha_{\rm def}$. The simulation data (symbols)
    are fitted by Eqs.\ \eqref{v_alpha} and
    \eqref{mean_run_length_alpha_defect}, respectively, with $q=0.25$
    for all curves.  (c) Run length histograms: Without stepping
    defects (the straight line), the run length distribution is well
    fitted by a single exponential with average run length as given by
    Eq. \eqref{mean_run_length}. With stepping defects, the
    distribution has a peak for short run lengths and decays faster
    for large run lengths. The motor parameters are chosen to match
    the traffic of kinesins
    \cite{Lipowsky_Nieuwenhuizen2001,Klumpp_Lipowsky2003}:
    $\alpha=0.0099333$, $\pi_{\rm ad}=1$, and $\epsilon=0.0001$.
    Furthermore we used $N_{\rm ch}=1960$ in (a) and (b) and
    $\rho_{\rm def}=0.001$ or $L/\ell\rm=1000$ and $N_{\rm ch}=316$ in
    (c).}
  \label{figure_single_motor_stepping_defects}
\end{figure}

If the stepping probability $\alpha_{\rm def}$ at the defect is not
large compared to the unbinding parameter $\epsilon$, the probability
for the motor to take the diffusion channel is comparable with the
probability to move forward along the filament. To estimate the
contribution of the diffusion channel, we start with the limiting case
$\alpha_{\rm def}=0$, for which the motor can only take the diffusive
channel to pass defect sites. We make the ansatz that the probability
for taking the diffusive pathway is proportional to the unbinding
probability $\epsilon$. For $\alpha_{\rm def}=0$, the time it takes
the motor to pass the defect is then given by $\tau_{\rm def}=\tau/q
\epsilon$ and the effective motor velocity is given by
\begin{equation}
  v_{\rm eff}\simeq\frac{L}{t_{\rm tot}} P_{\rm b}=
  \frac{\alpha}{1-\frac{\ell}{L}+\frac{\ell}{L}\frac{\alpha}{q \epsilon}}\ \frac{\pi_{\rm ad}}{\pi_{\rm ad}+\epsilon
    N_{\rm ch}}\ \frac{\ell}{\tau}. \label{v_0_alpha_defect}
\end{equation}
In these expressions, $q$ is an unknown free parameter and should
depend on the geometry of the system. For the parameters used in our
simulations, we have determined this parameter by fitting the
expression for $v_{\rm eff}$ to the the simulation data for
$\alpha_{\rm def}=0$, see Fig.
\ref{figure_single_motor_stepping_defects}(a), which leads to $q
\simeq 0.25$.

For intermediate values of the stepping probability $\alpha_{\rm
  def}$, both pathways contribute and the total probability to pass
the defect is given by the sum of the probabilities for the two
channels. The effective passing time $\tau_{\rm def}$ for the motor is
then proportional to $1/(\alpha _{\rm def}+q \epsilon)$.

This expression implies that the effective bound motor velocity (or
effective stepping rate) is given by
\begin{equation}
  v_{\rm b,eff}=\frac{\alpha}{1-\frac{\ell}{L}+\frac{\ell}{L}\frac{ 
      \alpha}{\alpha _{\rm def}+q \epsilon}}\ \frac{\ell}{\tau} \equiv \alpha_{\rm eff}\ \frac{\ell}{\tau}. \label{effective alpha}
\end{equation}
and leads to the effective motor velocity
\begin{equation}
  v_{\rm eff}= \frac{\alpha}{1-\frac{\ell}{L}+\frac{\ell}{L}\frac{ 
      \alpha}{\alpha _{\rm def}+q \epsilon}}
  \ \frac{\pi_{\rm ad}}{\pi_{\rm ad}+\epsilon N_{\rm ch}}\ \frac{\ell}{\tau},
  \label{v_alpha}
\end{equation}

For the parameters used in Fig.
\ref{figure_single_motor_stepping_defects}(a), the free parameter $q$
has been determined by fitting this expression to the simulation data
for $\alpha_{\rm def}=0$ and applied to other cases of different value
of $\alpha_{\rm def}$. The curves obtained from expression
\eqref{v_alpha} agree very well with the simulation data. We note
however that this expression is not strictly valid in the limit of
very weak defects with $\alpha_{\rm def}\approx\alpha$. Putting
$\alpha_{\rm def}=\alpha$ in \eq{effective alpha} leads to
$\alpha_{\rm eff}=\alpha+\mathcal{O}(\epsilon/\alpha)$, i.e. to a
discrepancy of order $\epsilon/\alpha$, which is very small for
processive motors.

Fig. \ref{figure_single_motor_stepping_defects}(a) shows that the
velocity is reduced compared to the case without defects. As one might
expect, this reduction is larger for stronger defects and/or for
higher defect densities. If the defects are sufficiently strong, even
very small defect densities lead to a substantial reduction of the
velocity. For example, if 1 percent of sites on the filament are
stepping defects with $\alpha_{\rm def}=0$, the velocity of the single
motor is reduced to about 20 percent compared to that without defects.


Another important property of the motors that is affected by stepping
defects is their run length $\Delta x$, i.e. the distance a motor
moves along a filament before unbinding from it. In the absence of
defect sites on the filament, the average run length is given by
\begin{equation}
  \langle\Delta x\rangle_0=\frac{\alpha}{4\epsilon/6}\ell \label{mean_run_length},
\end{equation}
and the distribution of run lengths decays exponentially as
$\exp(-\Delta x/\langle\Delta x\rangle_0)$. The run length
distribution for the case without defects is shown in Fig.
\ref{figure_single_motor_stepping_defects}(c), see the straight line.
This exponential decay is modified by the presence of stepping
defects, as also shown by the simulation data points in Fig.
\ref{figure_single_motor_stepping_defects}(c). In the presence of a
low density of defects, the run length distribution decays slightly
faster for large run length, and, in addition, develops a pronounced
peak for short run length.  This peak corresponds to short runs that
start close to a defect and end at the same defect. Both effects lead
to a reduction of the average run length. As the unbinding probability
of the motors is not affected by stepping defects, the time a motor
remains bound to a filament is the same with and without
defects.\footnote{We note that this is not necessarily the case if
  stepping and unbinding are coupled, e.g. if unbinding occurs during
  the step and is characterized by an unbinding probability per step
  rather than an unbinding rate. This scenario may arise for certain
  types of internal dynamics of the motors, see ref.
  \cite{Klumpp_Lipowsky2008}.} However, the distance moved during this
time is reduced if the motor encounters a defect. We can therefore
estimate the average run length using the effective velocity of a
bound motor, $v_{\rm b}$, which leads to
\begin{equation}
  \langle\Delta x\rangle=\frac{\alpha_{\rm eff}}{4\epsilon/6}\ell=\frac{\langle\Delta x\rangle_0}{1-\frac{\ell}{L}+\frac{\ell}{L}\frac{\alpha}{\alpha_{\rm def}+q\epsilon}}.  \label{mean_run_length_alpha_defect}
\end{equation} 
The dependence of the mean run length on the density and the strength
of the defects is shown in Fig.
\ref{figure_single_motor_stepping_defects}(b). For strong defects with
$\alpha_{\rm def}\ll\epsilon$, the precise value of $\alpha_{\rm def}$
is irrelevant, as motors at the defect site typically unbind, before
passing through the defect. For weaker defects, i.e. larger
$\alpha_{\rm def}$, the reduction of the run length is shifted towards
larger defect densities. For $\alpha_{\rm def}>\epsilon$, an
approximately two-fold reduction of the run length is obtained when
the defect density $\rho_{\rm def}$ and $\alpha_{\rm def}/\alpha$ have
the same order of magnitude.

\subsection{Many motors with stepping defects}
\label{Many motors with stepping defects}

We now consider the effect of stepping defects on the traffic of many
motors, which interact through mutual exclusion from filament sites.
For the traffic of many motors in a tube with length $L$, we are
interested in the following quantities \cite{Lipowsky_Klumpp2005}: (i)
Bound density $\rho_{\rm b}$ as a function of the spatial coordinate
$x$ along the tube axis; (ii) Bound current $J_{\rm b}(x)$ which gives
the number of motors that pass through a lattice site on the filament
with coordinate $x$ per unit time; and (iii) Average bound current
$\langle J_{\rm b} \rangle \equiv \int \,dx \rho_{\rm b}(x)/L$ which
characterizes the overall transport along the filament.

\begin{figure}
  \centering
  \includegraphics[width=12cm,clip]{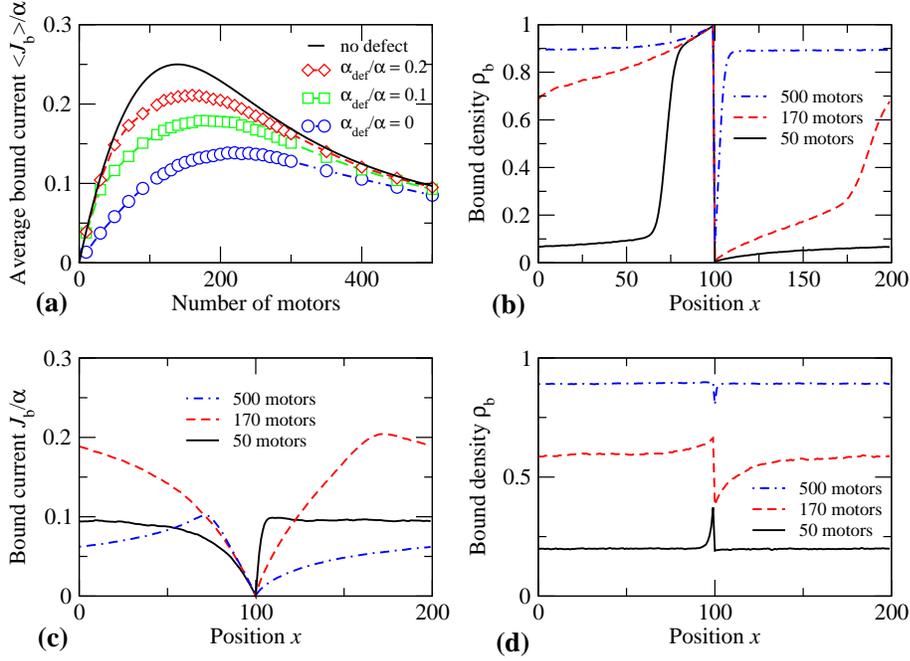}
  \caption{Traffic of molecular motors in the presence of stepping
    defects: (a) Normalized average bound current $\langle J_{\rm b}
    \rangle/\alpha$ (in units of $\tau^{-1}$) as a function of the
    number of motors in the tube for different values of $\alpha_{\rm
      def}$. The black curve which gives the exact value of average
    bound current for the case without defects is calculated as in
    Ref. \cite{Klumpp_Lipowsky2003}. (b) Bound motor density
    $\rho_{\rm b}$ as a function of the coordinate $x$ (in units of
    $\ell$) along the tube for $\alpha_{\rm def}=0$ and different
    numbers of motors in the tube; (c) Corresponding normalized bound
    motor current $J_{\rm b}/\alpha$ (in units of $\tau^{-1}$) as a
    function of $x$; and (d) Bound density profiles $\rho_{\rm b}(x)$
    for a weaker stepping defect with $\alpha_{\rm def}=0.5\alpha$.
    The parameters are the same as in
    \fig{figure_single_motor_stepping_defects} except for $L=200\ell$
    or, equivalently, $\rho_{\rm def}=0.005$.}
  \label{figure_many_motors_stepping_defects}
\end{figure}

In general, as one increases the total number of motors in the tube,
or equivalently the concentration of these motors, the average bound
current of motors on the filament increases first, but eventually
reaches a maximal value and then starts to decrease as the traffic
becomes jammed \cite{Klumpp_Lipowsky2003}, see Fig.
\ref{figure_many_motors_stepping_defects}(a). The presence of stepping
defects decreases the average bound current compared to the case
without the defects for all choices of the total number of motors
within the tube. The stronger the defects, the lower is the value of
the average bound current. In addition, the curve for the average
bound current as a function of the overall motor number becomes
broader as the strength of the defects increases, and the maximum of
the average bound current is shifted towards larger values of the
overall motor number, see Fig.
\ref{figure_many_motors_stepping_defects}(a).

Density profiles of bound motors along the filament in the presence of
stepping defects are shown in Fig.
\ref{figure_many_motors_stepping_defects}(b) for the limiting case
$\alpha_{\rm def}=0$, for which motors can only pass the defects by
unbinding, diffusion and rebinding to the filament. These profiles
show that stepping defects induce local traffic jams in front of the
defect and a depletion zone after it. These profiles are very similar
to those found in earlier studies on closed and half-open tube systems
\cite{Lipowsky_Nieuwenhuizen2001,
  Klumpp_Lipowsky2005,Mueller_Lipowsky2005} with the defect playing
the role of the boundary of the tube. The spatial extension of the
jammed region increases with the overall motor concentration. The end
of the jammed region distal to the defect is marked by a rather sharp
shock, i.e. a sudden change in density. The corresponding profiles of
the current of bound motors are shown in Fig.
\ref{figure_many_motors_stepping_defects}(c). Weaker stepping defects
with $\alpha_{\rm def}>0$ cause a smaller perturbation of the bound
density and bound current profiles; as shown in Fig.
\ref{figure_many_motors_stepping_defects}(d) for the case $\alpha_{\rm
  def}=0.5\alpha$, the effect of the defects is then confined to a
small region around the defect.


\section{Transport by molecular motors in the presence of unbinding
  defects}
\label{Transport by molecular motors in the presence of unbinding
  defects}


\subsection{Single motor with unbinding defects}

The second type of defects that we investigate is provided by
unbinding defects. Motors at an unbinding defect site move forward
with probability $\alpha$, unbind with probability $\epsilon_{\rm
  def}/ 6$ to each neighboring non-filament site and remain at the
same position with probability $\gamma_{\rm def} = 1 - \alpha -
4\epsilon_{\rm def} /6$. We study again the case where the defects are
regularly distributed on the filament with concentration $\rho_{\rm
  def}=1/L$ and we start by considering the effect of the defects on
the movement of a single motor.

Since all sites have the same stepping parameter $\alpha$, the
velocity of a bound motor, $v_\bd=\alpha\ell/\tau$ is not affected by
the unbinding defects and the effective velocity, which is averaged
over the bound and unbound states of the motor, is given by
\begin{eqnarray}
  v_{\rm eff}=\alpha   P_{\rm  b}\frac{\ell}{\tau}=\alpha   \left(\sum_{x} \rho_{\rm b} (x)\right) \frac{\ell}{\tau}.
  \label{current and P_b}
\end{eqnarray}

Since the unbinding defects break the translational invariance of the
system, they lead to inhomogeneous bound and unbound density profiles,
so that the binding-unbinding balance is again not valid locally. As
in the case of stepping defects, the bound and unbound motor densities
satisfy a global balance of binding and unbinding
\begin{equation}
  \pi_{\rm ad} \sum_{x}\rho_{\rm ub}(x,y_{\rm nn},z_{\rm nn})= \sum_{x} \epsilon(x)  \rho_{\rm b} (x),
  \label{balance_global_epsilon}
\end{equation}
where $y_{\rm nn}$ and $z_{\rm nn}$ are again the perpendicular
coordinates of a single channel of non-filament sites that are the
nearest neighbors of the filament sites.

As a global property of motor unbinding, we introduce an effective
unbinding probability, $\epsilon_{\rm eff}$, which is defined via
\begin{equation}
  \sum_x \epsilon(x)\rho_{\rm b}(x) \equiv \epsilon_{\rm eff} \sum_x \rho_{\rm b}(x)=\epsilon_{\rm eff} P_{\rm b}.
  \label{defination of effective unbinding probability}
\end{equation}
Using this relation in \eqref{balance_global_epsilon} together with
the replacement of $\rho_{\rm ub}(x,y_{\rm nn},z_{\rm nn})$ by
$\rho_{\rm ub}(x)$ and the normalization condition
\eqref{normalization condition for bound and unbound motors}, the
probability $P_{\rm b}$ as defined by Eq. \eqref{bound and unbound
  probabilities} is now given by
\begin{eqnarray}
  P_{\rm b}&=&\frac{\pi_{\rm ad}}{\pi_{\rm ad}+\epsilon_{\rm eff} N_{\rm ch}}
  \label{effective unbinding probability}
\end{eqnarray}
instead of relation \eqref{total bound probability for single motor
  with stepping defects} for stepping defects. Furthermore, the flux
balance relation \eqref{balance_global_epsilon} is equivalent to
\begin{equation}
  \pi_{\rm ad}\frac{P_{\rm ub}}{N_{\rm ch}}=\pi_{\rm ad}\frac{1-P_{\rm b}}{N_{\rm ch}}=\epsilon_{\rm eff}P_{\rm  b}. 
  \label{global balance with effective unbinding parameter}
\end{equation}

It follows from Eqs. \eqref{current and P_b} and \eqref{effective
  unbinding probability} which replaces the relation \eqref{total
  bound probability for single motor with stepping defects} for
stepping defects that the effective unbinding parameters
$\epsilon_{\rm eff}$ and the effective velocity $v_{\rm eff}$ satisfy
the relation:
\begin{equation}
  \epsilon_{\rm eff}=\frac{\pi_{\rm ad}}{N_{\rm ch}} \left( \frac{ 
      \alpha}{v_{\rm eff}}\frac{\ell}{\tau}-1\right).
  \label{epsilon_eff}
\end{equation}

So far, we have only rewritten the motor properties in terms of the
new parameter $\epsilon_{\rm eff}$. In the following, we will consider
several analytical approximations to determine the effective unbinding
parameter $\epsilon_{\rm eff}$, which then leads to estimates for the
bound state probability $P_\bd$ and the effective velocity $v_{\rm
  eff}$.

The simplest ansatz for $\epsilon_{\rm eff}$ is to take the average of
the unbinding probability along the filament, which leads to
\begin{equation}
  \epsilon_{\rm eff}\simeq \langle\epsilon(x)\rangle =\epsilon+\frac{\ell}{L}(\epsilon_{\rm def}-\epsilon)=\epsilon+\rho_{\rm def}(\epsilon_{\rm def}-\epsilon),
  \label{average_epsilon}
\end{equation}
with the defect density $\rho_{\rm def}=\ell/L$. This approximation is
valid if the bound motor density along the filament is approximately
constant. The latter condition is fulfilled if the motor is fast with
$\alpha\gg\epsilon$ and $\alpha\gg\epsilon_{\rm def}$.

\begin{figure}
  \centering
  \includegraphics[width=12cm,clip]{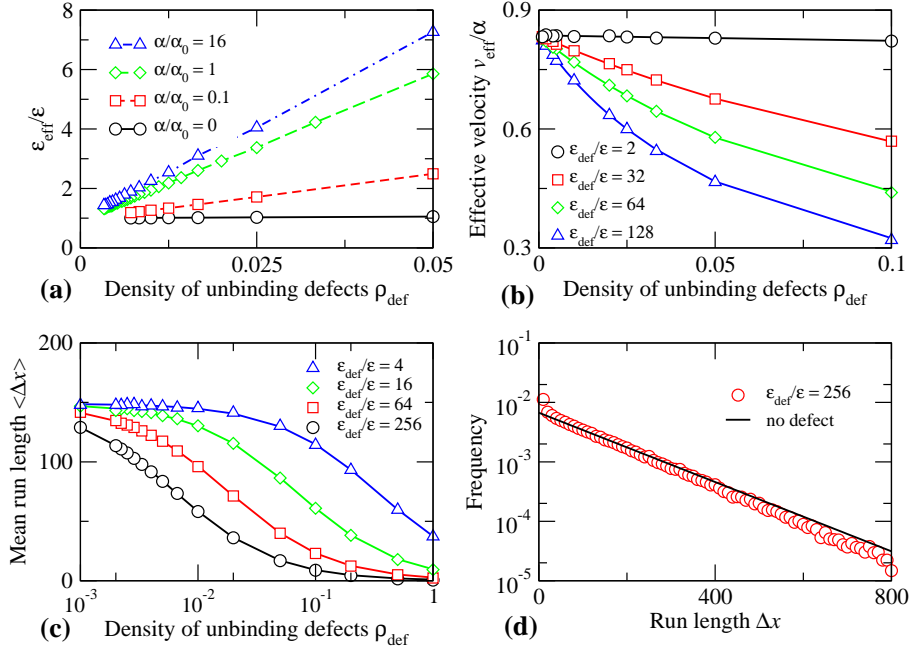}
  \caption{Transport properties of single motors in the presence of
    unbinding defects: (a) Normalized effective unbinding probability
    $\epsilon_{\rm eff}/\epsilon$ obtained from simulations via Eq.
    \eqref{epsilon_eff}; (b) Normalized effective velocity $v_{\rm
      eff}/\alpha$ (in units of $\ell/\tau$); and (c) Average run
    length $\langle\Delta x\rangle$ (in units of $\ell$) as a function
    of the defect density $\rho_{\rm def}$. The lines in (b) and (c)
    are fits to the simulation data using Eq.
    \eqref{epsilon_eff_fitting} combined with Eq. \eqref{epsilon_eff}
    and Eq.  \eqref{mean_run_length_epsilon_defect}, respectively; (d)
    Run length distribution for different defect strengths
    $\epsilon_{\rm def}/\epsilon$. The straight line corresponds to
    the exponential function $(4\epsilon/6\alpha\ell)\exp(- 4 \Delta x
    \epsilon/6\alpha\ell)$ of run length distribution without defects.
    The parameters are the same as in
    \fig{figure_single_motor_stepping_defects} except for $\alpha_0=0.0099333$ in (a) and $N_{\rm
      ch}=316$ and $L/\ell\rm=1000$ in (d).}
  \label{figure_single_motor_unbinding_defects}
\end{figure}

In the opposite limit of small stepping parameter $\alpha$, the flux
balance arising from binding and unbinding events is approximately
valid locally, i.e.
\begin{equation}
  \pi_{\rm ad}\rho_{\rm ub}(x) \approx \epsilon(x) \rho_{\rm b}(x)
  \label{local flux balance of binding and unbinding}
\end{equation}
for small $\alpha$. This relation is exact in the equilibrium case
with $\alpha=0$. Furthermore, for small $\alpha$, the unbound density
varies very little and can be approximated by a constant $\rho_{\rm
  ub}$, which again becomes exact for the equilibrium case with
$\alpha=0$. It then follows from \eqref{local flux balance of binding
  and unbinding} that the bound density $\rho_{\rm b}$ behaves as
\begin{equation}
  \rho_{\rm b}(x) \simeq \frac{\pi_{\rm ad}\rho_{\rm ub}}{\epsilon(x)},
\end{equation}
for small stepping probability $\alpha$. The probability $P_{\rm ub}$
for an unbound motor state is now given by
\begin{equation}
  P_{\rm ub}=N_{\rm ch}\sum_x \rho_{\rm ub}(x) \approx N_{\rm ch}\frac{L}{\ell}\rho_{\rm ub}
  \label{probability for unbound motor state with unbinding defects and small stepping parameter}
\end{equation}
and the probability for a bound motor state by
\begin{equation}
  P_{\rm b}=\sum_{x} \rho_{\rm b} (x) \approx \sum_{x} \frac{\pi_{\rm ad}\rho_{\rm ub}}{\epsilon(x)}. 
  \label{probability for bound motor state with unbinding defects and small stepping parameter}
\end{equation}
Inserting the two expressions \eqref{probability for unbound motor
  state with unbinding defects and small stepping parameter} and
\eqref{probability for bound motor state with unbinding defects and
  small stepping parameter} into Eq. \eqref{global balance with
  effective unbinding parameter}, one obtains the relation
\begin{equation}
  \frac{1}{\epsilon_{\rm eff}} \approx \frac{\ell}{L} \sum_{x=\ell}^{L}
  \frac{1}{\epsilon(x)} =\frac{1}{\epsilon}+
  \frac{\ell}{L}\left(\frac{1}{\epsilon_{\rm
        def}}-\frac{1}{\epsilon}\right)      .
  \label{average_epsilon_inverse}
\end{equation}
for the effective unbinding probability $\epsilon_{\rm eff}$ in the
limit of small stepping parameter $\alpha$. Note that in this limit of
small $\alpha$, one has to average the inverse of the local unbinding
parameter rather than the unbinding parameter itself as in the limit
of large $\alpha$. Expanding the relation
\eqref{average_epsilon_inverse} in powers of the defect density
$\rho_{\rm def}=\ell/L$ leads to
\begin{equation}
  \epsilon_{\rm eff} \approx \epsilon+\frac{\ell}{L}(\epsilon_{\rm def}-\epsilon)\frac{\epsilon}{\epsilon_{\rm def}}.
  \label{epsilon_eff_small_alpha}
\end{equation}
Comparison of this result for small $\alpha$ with Eq.
\eqref{average_epsilon}, which is valid for large $\alpha$, shows that
in both cases $(\epsilon_{\rm eff}-\epsilon)\sim \ell/L=\rho_{\rm
  def}$, but with different prefactors. These relations are confirmed
by simulations, see Fig.
\ref{figure_single_motor_unbinding_defects}(a). Furthermore, the
simulation data show how intermediate values of $\alpha$ interpolate
between these limiting cases. For these intermediate values, which are
typical for motors with finite processitivity, the effective unbinding
probability exhibits a weak dependence on $\alpha$. The simulation
data are described rather well by the expression
\begin{equation}
  \epsilon_{\rm eff} = \epsilon+\rho_{\rm def}(\epsilon_{\rm def}-\epsilon)\frac{\epsilon+q^{\prime}\alpha}{\epsilon_{\rm def}+q^{\prime}\alpha},
  \label{epsilon_eff_fitting}
\end{equation}
where $q^{\prime}$ is a free parameter determined to be $q^{\prime}
\simeq 4.0$ by fitting the simulation data. Note that the expression
\eqref{epsilon_eff_fitting} interpolates between Eq.
\eqref{average_epsilon} for large $\alpha$ and Eq.
\eqref{epsilon_eff_small_alpha} for small $\alpha$. Using Eq.
\eqref{epsilon_eff_fitting} also leads to a rather accurate
description of the motor velocity as obtained from simulations, see
Fig. \ref{figure_single_motor_unbinding_defects}(b), where the motor
velocity is shown as a function of defect density and defect strength,
i.e. the defect unbinding probability. We note that small defect
densities can have a rather strong effect, if the unbinding
probability at the defect is of the same order of magnitude as the
stepping probability $\alpha$. For example in the case $\epsilon_{\rm
  def}=128\epsilon$, a defect density of about 4 percent reduces the
effective velocity two-fold and a 10 percent defect density reduces it
three-fold.

Unbinding defects also reduce the run length of the motor. In the
presence of unbinding defects, the mean run length can be expressed in
terms of the effective unbinding probability as
\begin{equation}
  \langle\Delta x\rangle=\frac{\alpha}{4\epsilon_{\rm eff}/6}\ell,
  \label{mean_run_length_epsilon_defect}
\end{equation}
as obtained from Eq. \eqref{mean_run_length} when $\epsilon$ is
replaced by $\epsilon_{\rm eff}$. The dependence of the mean run
length on the defect strength and the defect density is shown in Fig.
\ref{figure_single_motor_unbinding_defects}(c). As in the case of
stepping defects studied above, see
\fig{figure_single_motor_stepping_defects}, small defect densities
have only a weak effect on the run length, while large defect
densities shorten the runs strongly. The crossover density at which
the effect of the defects becomes notable has a strong dependence on
the defects strength, i.e. $\epsilon_{\rm def}$, and can be quite
small for strong defects with large $\epsilon_{\rm def}$. For example,
for $\epsilon_{\rm def}=256\epsilon$, a two-fold reduction of the mean
run length is obtained for $\rho_{\rm def}\simeq 0.007$, that is if
less than one percent of the filament sites are unbinding defects. The
effect of unbinding defects on the run length is very similar to the
corresponding effect of stepping defects, see
\fig{figure_single_motor_stepping_defects}. This similarity reflects
the fact that the effects of strong stepping and unbinding defects
have some common aspects: when the motor encounters the defects, it
has a rather high probability to unbind from the filament, either
because of the high unbinding probability at an unbinding defect or
because of the prolonged sojourn time at a stepping defect.

As a consequence, unbinding and stepping defects also have a similar
effect on the run length distributions, compare Fig.
\ref{figure_single_motor_unbinding_defects}(d) and
\ref{figure_single_motor_stepping_defects}(c). Fig.
\ref{figure_single_motor_unbinding_defects}(d) shows run length
distributions for a rather low density of unbinding defects. As in the
case of stepping defects, the length scale that governs the
exponential decay of the distribution is slightly reduced, and the
distributions exhibit a peak at small run lengths.

\subsection{Many motors with unbinding defects }

\begin{figure}
  \centering
  \includegraphics[width=12cm,clip]{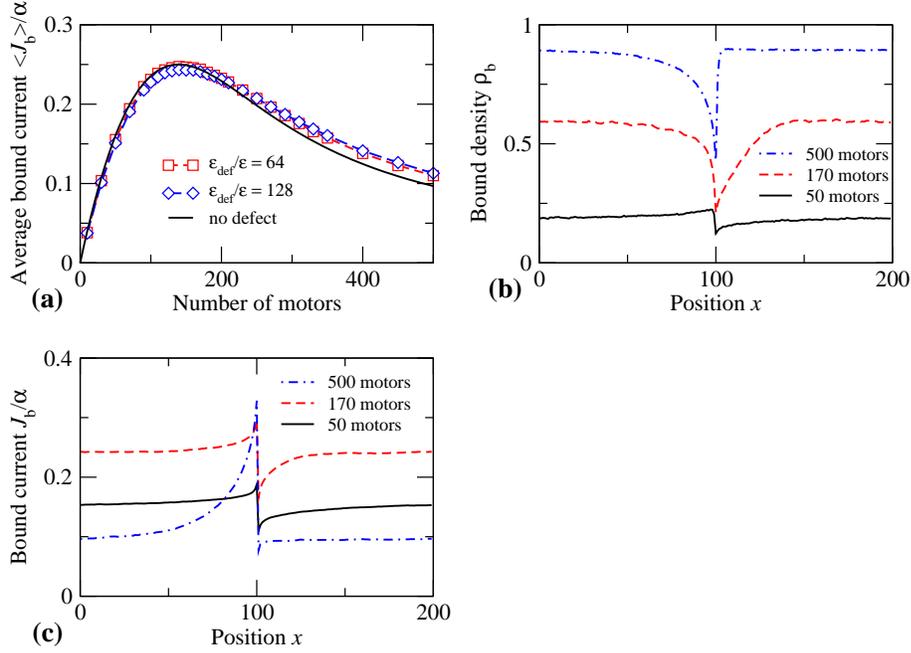}
  \caption{Motor traffic in the presence of unbinding defects: (a)
    Normalized average bound current $\langle J_{\rm b}
    \rangle/\alpha$ (in units of $\tau^{-1}$) as a function of the
    number of motors for different values of $\epsilon_{\rm def}$. The
    black curve which gives the exact value of average bound current
    for the case without defects is calculated as in Ref.
    \cite{Klumpp_Lipowsky2003}. (b) Bound motor density $\rho_{\rm b}$
    as a function of the spatial coordinate $x$ (in units of $\ell$)
    along the tube axis; and (c) Corresponding profiles of the bound
    motor current $J_{\rm b}(x)$ (in units of $\tau^{-1}$). The
    parameters are the same as in
    \fig{figure_many_motors_stepping_defects}, with $\epsilon_{\rm
      def}=128\epsilon$ in (b) and (c).}
  \label{figure_many_motors_unbinding_defects}
\end{figure}

Now let us consider the effect of unbinding defects on the traffic of
many motors which interact through mutual exclusion. Fig.
\ref{figure_many_motors_unbinding_defects}(a) shows the average bound
current as a function of the overall number of motors within the tube
for a low defect density. It can be seen from these plots that the binding defects do not always reduce the current, as one might 
expect and as found for the  stepping 
defects  discussed above, see Fig.\ \ref{figure_many_motors_stepping_defects}(a). 
For small motor numbers, the average bound
current is indeed slightly reduced by the presence of the defects, but for
large motor numbers, the current is slightly increased. This observation can be
understood as follows: For small motor concentration, the decrease of
the bound motor density arising from the unbinding defects leads to a
reduction of the average bound current. If the concentration of motors
is however larger than the concentration for which the average bound
current attains its maximal value, a reduction of the bound motor
density leads to an increase of the average bound current, because
the increased unbinding probability relieves the traffic jams
appearing for high motor densities.

Profiles of the bound motor densities on the filament are shown in
Fig. \ref{figure_many_motors_unbinding_defects}(b). The profiles are
rather constant away from the defect, but have a minimum at the defect
site. This is what one would expect, since motors unbind from the
filament at this site. For small overall motor concentration (or total
motor number), the profiles also exhibit a maximum in front of the
defect. This maximum arises from the locally increased density of
unbound motors which leads to increased rebinding of motors to the
filament. Since this maximum requires a locally increased motor
density, it is only present if the diffusion of unbound motors is not
too fast. When the overall motor concentration is increased, this
maximum disappears. The corresponding bound current profiles are shown
in Fig. \ref{figure_many_motors_unbinding_defects}(c) for a strong
defect with $\epsilon_{\rm def}=128\epsilon$. The current exhibits a
peak in front of the defects and a depletion zone behind these
defects. The depletion zone behind the defect follows the bound density profile closely, which indicates that it reflects the reduced motor density due to unbinding at the defect. On 
the other hand, the peak in front of the defect is present when the bound density 
exhibits a peak, as well as for high motor concentration when no density peak occurs.


\section{Transport by molecular motors in the presence of binding
  defects}
\label{Transport by molecular motors in the presence of binding
  defects}


\subsection{Single motor with binding defects}

The third class of defects we investigate are binding defects. This type of defect appears to be the most common one in biological systems as shown in Table \ref{Examples of defects and their effect on motor properties}. At a binding defect site, bound
motors have the same hopping probabilities
as at any other filament site, but unbound motors that approach 
the binding defect site have a reduced sticking probability
$\pi_{\rm def}$.

Similar to the case of unbinding defects as studied in the previous
section, binding defects do not affect the movement of bound motors,
but rather change the balance of binding and unbinding. The stronger
the binding defects are, or the higher the density of binding defects
on the filament is, the less likely it becomes for motors to bind to
the filament. A local balance of binding and unbinding, similar to Eq.
\eqref{balance_global_epsilon}, is also valid in this case, but now
with a site-dependent binding probability.

\begin{figure}
  \centering
  \includegraphics[width=12cm,clip]{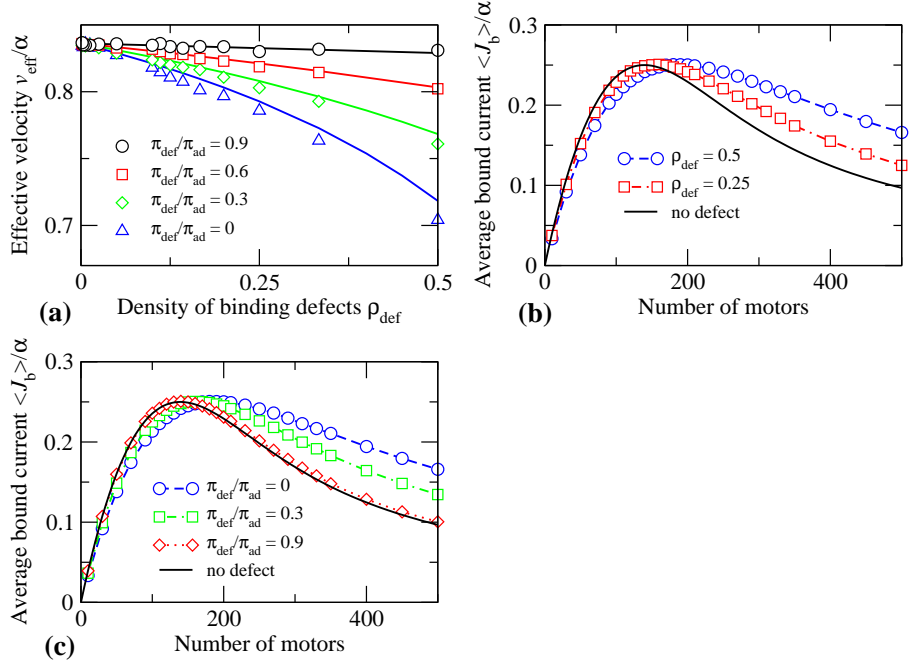}
  \caption{Transport properties of single motors and motor traffic in
    the presence of binding defects: (a) Normalized velocity $v_{\rm
      eff}/\alpha$ of a single motor as a function of the density
    $\rho_{\rm def}$ of binding defects for different values of
    $\pi_{\rm def}$. The symbols show simulation data and the lines 
    the corrsponding results from a mean-field calculation; 
    (b) Normalized average bound current $\langle
    J_{\rm b} \rangle/\alpha$ (in units of $\tau^{-1}$) as a function
    of the number of motors in the tube for $\pi_{\rm def}=0$ and
    different defect densities $\rho_{\rm def}$; and (c) Motor current
    (in units of $\tau^{-1}$) for $\rho_{\rm def}=0.5$ and different
    defects strengths $\pi_{\rm def}$. The black curves in (b) and (c)
    which give the exact value of average bound current for the case
    without defects are calculated as in Ref.
    \cite{Klumpp_Lipowsky2003}. The other parameters are the same as
    in \fig{figure_many_motors_stepping_defects}.}
  \label{figure_binding_defects}
\end{figure}

Simulation results for the effective motor velocity as a function of
the density of binding defects are shown in Fig.
\ref{figure_binding_defects}(a). This figure also includes results obtained from a mean field approximation using an effective (site-independent) binding parameter $\pi_{\rm eff} \simeq \langle
\pi_{\rm ad}(x) \rangle=\pi_{\rm ad}+\rho_{\rm def}(\pi_{\rm def}-
\pi_{\rm ad})$. As can be seen in Fig.
\ref{figure_binding_defects}(a), the latter approximation leads to good agreement with the simulation data. The most noticeable feature of Fig.
\ref{figure_binding_defects}(a) is that the effect of binding defects is rather weak. Binding defects only have a
notable effect  at high defect densities. But even if binding is
completely suppressed at every second filament site, i.e. for
$\rho_{\rm def}=0.5$ and $\pi_{\rm def}=0$, the effect remains weak, 
since the effective motor velocity is only
decreased by 14 percent. This result is in striking contrast to the
strong effects of the other two defects types.

As the binding defects do not affect the movement of bound motors, the
run length is not changed compared to the case without defects.

\subsection{Many motors with binding defects}
\label{Many motors with binding defects}

Finally we investigate the traffic of many motors in the presence of
binding defects. Fig.  \ref{figure_binding_defects}(b) shows the
average bound current as a function of the number of motors in the
tube for binding defects with $\pi_{\rm def}=0$. As in the case of unbinding defects shown in Fig.\ \ref{figure_many_motors_unbinding_defects}(a), binding defects can both increase and decrease the bound motor current. For low motor
concentrations, binding defects reduce the current by reducing the
probability that a motor is bound. For large motor concentrations, the binding
defects increase the current compared to the case without defects,
because the binding defects reduce the motor traffic jams on the
filament. For weaker binding defects with $\pi_{\rm def}>0$, the
effect is similar, but even smaller. The maximal value of the average
bound current is not changed by the presence of binding defects, the
defects rather shift the current maximum to larger motor numbers, see
Fig. \ref{figure_binding_defects}(b) and
\ref{figure_binding_defects}(c). Again, the effect of binding defects
is much weaker than that of stepping or unbinding defects. Binding
defects only have a notable effect on the traffic of many motors when
the defect density is sufficiently large.

\subsection{Binding defects and cooperative transport by several motors}
\label{Cargo_severalMotors}

We have noted above that binding defects account for most of the biologically relevant defects. In striking contrast to their importance in biological systems, our analysis shows that binding defects have very small effects on both the movements of individual motors and on the traffic of many motors. It is important to note, however, that our conclusions about binding
defects apply only to the traffic of individual motor molecules or to
the traffic of cargo particles that are pulled by single motors.
Unbinding defects are expected to have a much stronger effect on cargo
particles that are pulled by teams of several motors, the typical
situation for {\it in vivo} transport
\cite{Klumpp_LipowskyCoopTr,Beeg__Lipowsky2008,Mueller_Lipowsky2008}.
Thus, let us consider a cargo particle that is pulled by $N$ identical
motors such as kinesin. The effective unbinding rate ${\overline
  \epsilon}_{\rm eff}$ of such a cargo particle is proportional to
$(\epsilon/\pi_{\rm ad})^{N - 1}$ for strongly binding motors with
$\epsilon/\pi_{\rm ad} \ll 1$ \cite{Klumpp_LipowskyCoopTr} where
$\epsilon$ and $\pi_{\rm ad}$ are the previously defined unbinding
and sticking probabilities of a single motor. Thus, for a cargo
particle pulled by $N$ strongly binding motors, the effective
unbinding rate ${\overline \epsilon}_{\rm eff} \sim (1/\pi_{\rm
  ad})^{N-1}$ and is, thus, strongly affected by the value $\pi_{\rm
  ad}$ for the sticking probability of a single motor. This implies
that {\em binding} defects for single motors will act as {\em
  unbinding} defects for cargo particle pulled by several motors. This effect has indeed been demonstrated experimentally for tau proteins: As expected for binding defects,  tau proteins do not affect the run length of individual motors \cite{Seitz_Mandelkow_2002,Vershinin_Gross2007}. However, tau proteins strongly reduce the run length for cargoes pulled by several motors \cite{Vershinin_Gross2007} and, thus, act as effective  unbinding defects. A
quantitative description for this latter effect can be obtained by an
extension of the models studied here and in Ref.
\cite{Klumpp_LipowskyCoopTr,Mueller_Lipowsky2008}.


\section{Summary}
\label{Summary}


In this article, we have studied the traffic of molecular motors in
the presence of different types of static defects on the filament. We
have determined several properties that characterize the movement of
single motors as well as the traffic behavior in many-motor systems
such as motor velocities, motor run length, motor density and current
profiles.

We have considered three basic types of static defects, namely
stepping, unbinding and binding defects. At the defect sites, the
dynamics of the motors differs in only one transition probability from
the dynamics at other filament sites. While stepping defects and
unbinding defects have rather strong effects on the motor behavior and
severely reduce the velocity, run length and currents, the effect of
binding defects on individual motors is much weaker and becomes only
notable if the density of the defects is sufficiently large. The run
length is not affected at all by binding defects. At first sight,
these results appear to be at odds with the experimental observation
that most biologically relevant defects such as MAPs represent binding defects as summarized in Table
\ref{Examples of defects and their effect on motor properties}. It is,
however, plausible that MAPs mainly regulate the movement of larger
cargoes, which are pulled by several motors. For such cargo particles,
the effective unbinding rate $\overline{\epsilon}_{\rm eff}$ depends
rather strongly on the binding probability $\pi_{\rm ad}$ of
individual motors as discussed in subsection \ref{Cargo_severalMotors}. 
Thus, in order to describe the regulation of cargo
particles by, e.g., MAPs, one should extend the models discussed here
to cooperative transport by teams of motors.

In general, localized inhomogeneities on filaments, which may be both
modifications of the filament itself or other molecules bound to the
filaments, can modulate the patterns of molecular motor transport in
various ways. While most of the systems we considered reduce motor
movements, we note that unbinding defects can increase the motor
current if the local motor density is high. In addition, it is easy to
imagine binding defects that {\em increase} motor binding and function
as loading stations that initiate filament transport, although we are
not aware of any biological system with this function. In addition to
their functions for intracellular transport, regulatory mechanisms via
filament defects or inhomogeneities may also be of interest for the
development of artificial biomimetic transport systems based on
molecular motors
\cite{Hess_Bachand_Vogel2004,vandenHeuvel_Dekker2007}.




\begin{acknowledgements}
  The authors thank Melanie M{\"u}ller for fruitful discussions. SK
  was supported by Deutsche Forschungsgemeinschaft (Grants KL818/1-1
  and 1-2) and by the NSF through the Center for Theoretical
  Biological Physics (Grant PHY-0822283).
\end{acknowledgements}



\end{document}